\documentclass[iop]{emulateapj}

\usepackage{newtxtext,newtxmath}

\usepackage[T1]{fontenc}
\usepackage{ae,aecompl}

\usepackage{graphicx}	
\usepackage{amsmath}	

\usepackage{amssymb}	
\usepackage{bm}
\usepackage{booktabs}
\usepackage{cleveref}
\usepackage{natbib}

\newcommand{\secpoint}{\mbox{$''\mskip-7.6mu.\,$}}
\newcommand{\angstrom}{\mbox{\normalfont\AA}}

\newcommand{\fesc}{$f_{\rm esc,HII}$ }

\newcommand{\pasa}{PASA}
\newcommand{\jwst}{ {\it JWST}}
\newcommand{\hst}{{ \it HST}}

\newcommand{\rmxaa}{RMXAA}
\begin{document}

\title{Searching for Extremely Blue UV Continuum Slopes at $\MakeLowercase{z}=7-11$ in JWST/NIRCam Imaging: Implications for Stellar Metallicity and Ionizing Photon Escape in Early Galaxies}

\author{
Michael W. Topping\altaffilmark{1},
Daniel P. Stark\altaffilmark{1},
Ryan Endsley\altaffilmark{1},
Adele Plat\altaffilmark{1},
Lily Whitler\altaffilmark{1},
Zuyi Chen\altaffilmark{1},
St\'ephane Charlot\altaffilmark{2}
}

\altaffiltext{1}{Steward Observatory, University of Arizona, 933 N Cherry Ave, Tucson, AZ 85721, USA}
\altaffiltext{2}{Sorbonne Universit\'e, UPMC-CNRS, UMR7095, Institut d'Astrophysique de Paris, F-75014, Paris, France}

\email{michaeltopping@arizona.edu}

\shortauthors{Topping et al.}

\shorttitle{Blue UV Slopes at $z>7$}

\begin{abstract}
The ultraviolet (UV) continuum slope ($\beta~$where$~f_\lambda\propto\lambda^\beta$) of galaxies is sensitive to a variety of properties, from the metallicity and age of the stellar population to dust attenuation throughout the galaxy. Considerable attention has focused on identifying reionization-era galaxies with very blue UV slopes ($\beta<-3$). Not only do such systems provide a signpost of low-metallicity stars, but they also identify galaxies likely to leak ionizing photons from their HII regions as such blue UV slopes require the reddening effect of nebular continuum to be diminished. In this paper we present a search for reionization-era galaxies with very blue UV colors in recent JWST/NIRCam imaging of the EGS field. We characterize UV slopes for a large sample of$~z\simeq7-11~$galaxies, finding a median of$~\beta=-2.0$. Two lower luminosity (M$_{\rm{UV}}\simeq-19.5$) and lower stellar mass (6-10$\times10^7$M$_\odot$) systems exhibit extremely blue UV slopes ($\beta=-2.9~$to$~-3.1$) and rest-optical photometry indicating weak nebular line emission. Each system is very compact (r$_e\lesssim$260pc) with very high star-formation-rate surface densities. We model the SEDs with a suite of BEAGLE models with varying levels of ionizing photon escape. The SEDs cannot be reproduced with our fiducial (f$_{\rm{esc,HII}}$=0) or alpha-enhanced (Z$_\star<Z_{\rm{ISM}}$) models. The combined blue UV slopes and weak nebular emission are best-fit by models with significant ionizing photon escape from HII regions (f$_{\rm{esc,HII}}$=0.5-0.8) and extremely low-metallicity massive stars (Z$_\star$=0.01-0.06Z$_\odot$). The discovery of these galaxies highlights the potential for JWST to identify large numbers of candidate Lyman Continuum leaking galaxies in the reionization-era and suggests low-metallicity stellar populations may be common in dwarf galaxies at $z>7$.

\end{abstract}


%
%
\section{Introduction}
\label{sec:intro}
Deep multi-wavelength imaging surveys have been at the forefront of studies of galaxy formation at high redshift.  The sensitivity and large sample sizes comprising such surveys have utilized models of the integrated spectral energy distributions (SEDs) produced by galaxies to infer many key properties of the galaxy population, including stellar masses, stellar population ages, star-formation histories, and strengths of strong nebular emission lines \citep[e.g.,][]{Ono2012, Labbe2013, Smit2014, Bouwens2014, Bouwens2015, Finkelstein2015, Roberts-Borsani2016, Stark2016, deBarros2019, Endsley2021, Stefanon2022}. New datasets from the {\it James Webb Space Telescope } (\jwst) will soon  dramatically advance the available sample sizes and measurement precision probing galaxies in the earliest epochs.

The power-law slope of the ultraviolet (UV) continuum ($\beta$ where f$_\lambda\propto \lambda^\beta$) has been a prominent metric for establishing galaxy properties in the high-redshift Universe. Great strides were made in the construction of statistical samples at high redshift brought on by near-infrared imaging using the \hst/WFC3 camera. The derived UV continuum slopes are found to be blue, with values of roughly $\beta=-2$ being common at $z\simeq 7$ \citep[e.g.,][]{McLure2011, Finkelstein2012b, Dunlop2012, Rogers2013, Bouwens2014,Bhatawdekar2021}. The UV slopes have been shown to become bluer both at lower UV luminosities and at higher redshifts \citep[e.g.,][]{Bouwens2014},  suggesting the $z\simeq 7$ population likely faces less attenuation from dust than is common in galaxies at later times.

Early efforts with {\it HST} were driven by the search for the bluest galaxies, with the motivation that these may provide a signpost of stellar populations dominated by very massive stars and extremely low metallicities. A variety of programs have presented measurements claiming a small population of faint galaxies at $z\gtrsim7$ exhibiting very blue UV slopes ($\beta\simeq -3$) \citep[e.g.,][]{Bouwens2010, Labbe2010,Ono2010,Jiang2020}.  While such 
measurements face considerable uncertainty owing to the 
underlying photometric errors, the discovery of these sources  have nonetheless pushed the limits of stellar population models (see \citealt{Bouwens2010, Wilkins2011}). Many models do suggest that UV continuum slopes approaching $\beta\simeq -3$ can in principle be 
produced in young and low metallicity stellar populations, but in practice they are not expected to be observed. 
At the young stellar population ages where such blue slopes 
are found, nebular emission 
from two-photon and free-bound continuum processes (hereafter nebular continuum) significantly redden the UV continuum slopes \citep[e.g.,][]{Nussbaumer1984,Bottorff2006}. 

The only way to achieve 
UV slopes approaching values of $\beta\simeq -3$ is if the 
nebular continuum contribution is significantly reduced, as would occur if ionizing photons escape from HII regions without being reprocessed into nebular emission \citep{Bouwens2010, Ono2010, Raiter2010b, Robertson2010}. 
While escape of ionizing photons from HII regions is only the first step toward leakage (the radiation also must escape the ISM and CGM), it is a necessary precondition and sources with a large fraction of their ionizing radiation escaping from HII regions (hereafter \fesc) should be excellent Lyman continuum (LyC) leaker candidates.  As such, 
the detection of galaxies with $\beta\simeq -3$ may identify extremely efficient ionizing agents, with hard spectra from low metallicity stars and large escape fractions.

The detections described above have motivated attempts to 
develop the modeling infrastructure necessary to link 
very blue UV slopes to information on ionizing photon 
escape fractions. Early papers explored implications for 
some of the first detected sources with $\beta < -2.8$ \citep{Ono2010, Bouwens2010, Wilkins2011}, but constraints on \fesc were quite limited given the available data. In more recent studies, efforts have focused on modeling the reduction in strength of nebular emission lines that 
arises when there is significant ionizing photon escape 
(e.g., \citealt{Robertson2010,Zackrisson2013,Zackrisson2017,Plat2019}).
\citet{Zackrisson2013} motivate that both UV slope and 
H$\beta$ EW can be used simultaneously to identify 
galaxies with significant LyC leakage. Meanwhile, 
observations have begun to develop large samples of 
LyC leakers at lower redshifts (e.g. \citealt{Flury2022a,Fletcher2019,Pahl2022}), allowing these models to be more directly tested (e.g., \citealt{Yamanaka2020,Flury2022b,Chisholm2022}). 
And with the commencement of {\it JWST} Cycle 1, it is rapidly becoming possible to better constrain the extremely blue tail of the distribution of UV slopes at $z>7$. And 
unlike in the past, it is now possible to provide robust constraints on the rest-optical emission line strengths of $z>7$ objects with blue colors, allowing much-improved modeling of the influence of ionizing photon escape on the SED.

In this paper, we present results stemming from a search for galaxies with very blue UV continuum slopes among the $z>7$ population in the recent NIRCam imaging of the EGS field by {\it JWST}.  We identify two galaxies with both extremely blue UV slopes and rest-optical photometry that is suggestive of relatively weak 
nebular lines, both indicative of LyC leakage and low metallicity stellar populations. We fit the SEDs with a 
suite of models allowing variation of \fesc \citep{Plat2019} and discuss the properties of the sources in the context of 
known LyC leakers at lower redshifts (e.g., \citealt{Flury2022a,Pahl2022}).
Section 2 provides a summary of the models that we use in this paper. Section 3 briefly introduces our data products and reduction techniques (\S3.1), before discussing our search for very blue galaxies in $z\simeq 7-11$ galaxy samples in NIRCam imaging. In Section 4, we investigate these extremely blue objects in the context of models that include the effects of ionizing radiation escape and compare the properties of our candidates to known LyC leakers at lower redshifts. Finally, Section 5 presents a brief summary.  Throughout this paper we use AB magnitudes \citep{Oke1984}, and assume a cosmology with $\Omega_m = 0.3$, $\Omega_{\Lambda}=0.7$, $H_0=70 \textrm{km s}^{-1}\ \textrm{Mpc}^{-1}$. We additionally adopt solar abundances from \citet[][i.e., $Z_{\odot}=0.014$]{Asplund2009}.

\begin{figure*}
    \centering
    \includegraphics[width=1.0\linewidth]{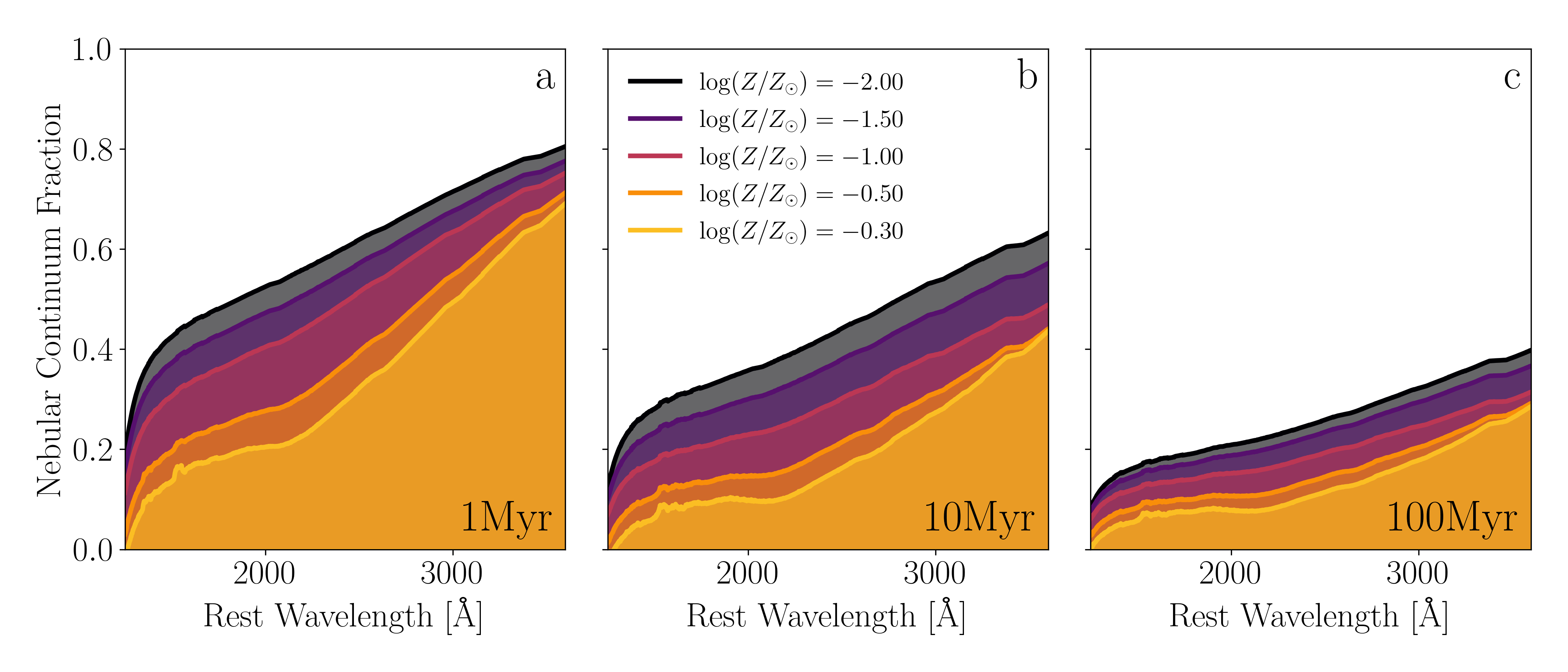}
    \caption{Fraction of the total flux that is the nebular continuum as a function of rest wavelength. We show the results for models of 1Myr, 10Myr, and 100Myr in panels a, b, and c, respectively.  In each panel we display continuum fractions calculated for five different assumed metallicities.  The nebular continuum is strongest at the youngest ages and lowest metallicities, and decreases in strength toward older ages and higher metallicities.}
    \label{fig:nebularfrac}
\end{figure*}

%
%
\section{Stellar Population Models}
\label{sec:uvslopes}

In this paper, we investigate the presence of extremely blue UV slopes in star forming galaxies at $z>7$.  Our goals in doing so are both to characterize the stellar  populations in the reionization era and to search for galaxies that may be  Lyman Continuum leaking candidates. To interpret these systems, we  will compare the observed SEDs to a suite of models that allow for a range of escape fractions using density bounded HII regions \citep{Plat2019}. The models make use of the BayEsian Analysis of GaLaxy sEds tool \citep[BEAGLE;][]{Chevallard2016}  (described in more detail below) and utilize the latest version of the \citet{cb2003} stellar population synthesis models. BEAGLE self-consistently includes the contribution of nebular emission calculated using the photoionization models described in \citet{Gutkin2016}. The density bounded models that form the 
basis of our varying escape fraction models have been developed and described in detail in \citet{Plat2019}. In this section, we discuss the UV slopes expected in our fiducial ionization-bounded models (\fesc = 0) and then describe how both the UV slopes and rest-optical emission 
lines vary as escape fraction is increased.

The sensitivity of $z>7$ SEDs to the presence of ionizing photon escape is a direct result 
of the prominence of nebular continuum emission in the SEDs. The shift of the galaxy population 
toward larger specific star formation rates (sSFRs) at fixed mass between $z\simeq 2$ 
and $z\simeq 7$ \citep{Topping2022} implies a corresponding shift toward younger 
light-weighted ages \citep{Whitler2022}.  With the considerable effort put into characterization of SEDs with {\it HST} and {\it Spitzer} in the last two decades, we now have a reasonably 
good idea of the distribution of light-weighted ages up to $z\simeq 7$, with significant evidence 
that a substantial fraction of the population is dominated by the light from very young 
stellar populations \citep{Smit2014,Endsley2021,Stefanon2022}.  At these young ages, 
nebular continuum emission makes a significant contribution to the total continuum flux 
from the far-UV into the optical.

In Figure~\ref{fig:nebularfrac} we show the fraction of the observed continuum that is of nebular origin as a function of  wavelength for our fiducial ionization-bounded BEAGLE 
models. We show the nebular continuum fraction for three different ages, representative of the range of ages seen at $z\simeq 7$ and we highlight the dependence on metallicity.
These have been run with our standard parameterization that assumes a constant star-formation history, a \citet{Chabrier2003} IMF with an upper-mass cutoff of $300M_{\odot}$, $\log(U)=-2.5$, a solar abundance pattern, and no dust. 
At young ages, the nebular continuum contributes a sizable fraction of the total flux at all wavelengths.  At low metallicities, galaxies with an age of 1 Myr have SEDs that are composed of $\sim40\%$ nebular continuum at 1500\AA, rising to nearly  $\sim80\%$ at the blue side of the Balmer jump. While the normalization of this continuum fraction decreases with age, at 10Myr the contribution is still over a third of the total flux, and comprises 10-20\% at 100 Myr. Clearly, this represents a significant effect in galaxies that are common at $z>7$. This figure additionally 
illustrates the significant effect that the metallicity has on the nebular contribution. The lower metallicity models  (with their associated higher electron temperatures) yield SEDs with a more significant contribution of the nebular continuum at all wavelengths.  

It is clear from Figure~\ref{fig:nebularfrac} that the contribution of nebular continuum is more prominent in the near-UV and optical. We thus may expect to see very blue UV slopes in the far-UV (where stellar continuum contributes over 50\% of the light) with slightly redder slopes in near-UV where the contribution of nebular continuum is greater. As a result, it will be important to consider the UV slopes in 
both the FUV and NUV if interested in whether nebular continuum is making a significant contribution. Objects with significant leakage (and hence diminished 
nebular continuum) would be expected to show blue UV slopes in both the FUV and 
NUV, whereas those with very young SEDs and minimal leakage may show a break in their 
continuum slopes between the FUV and NUV.

\begin{figure*}
    \centering
    \includegraphics[width=1.0\linewidth]{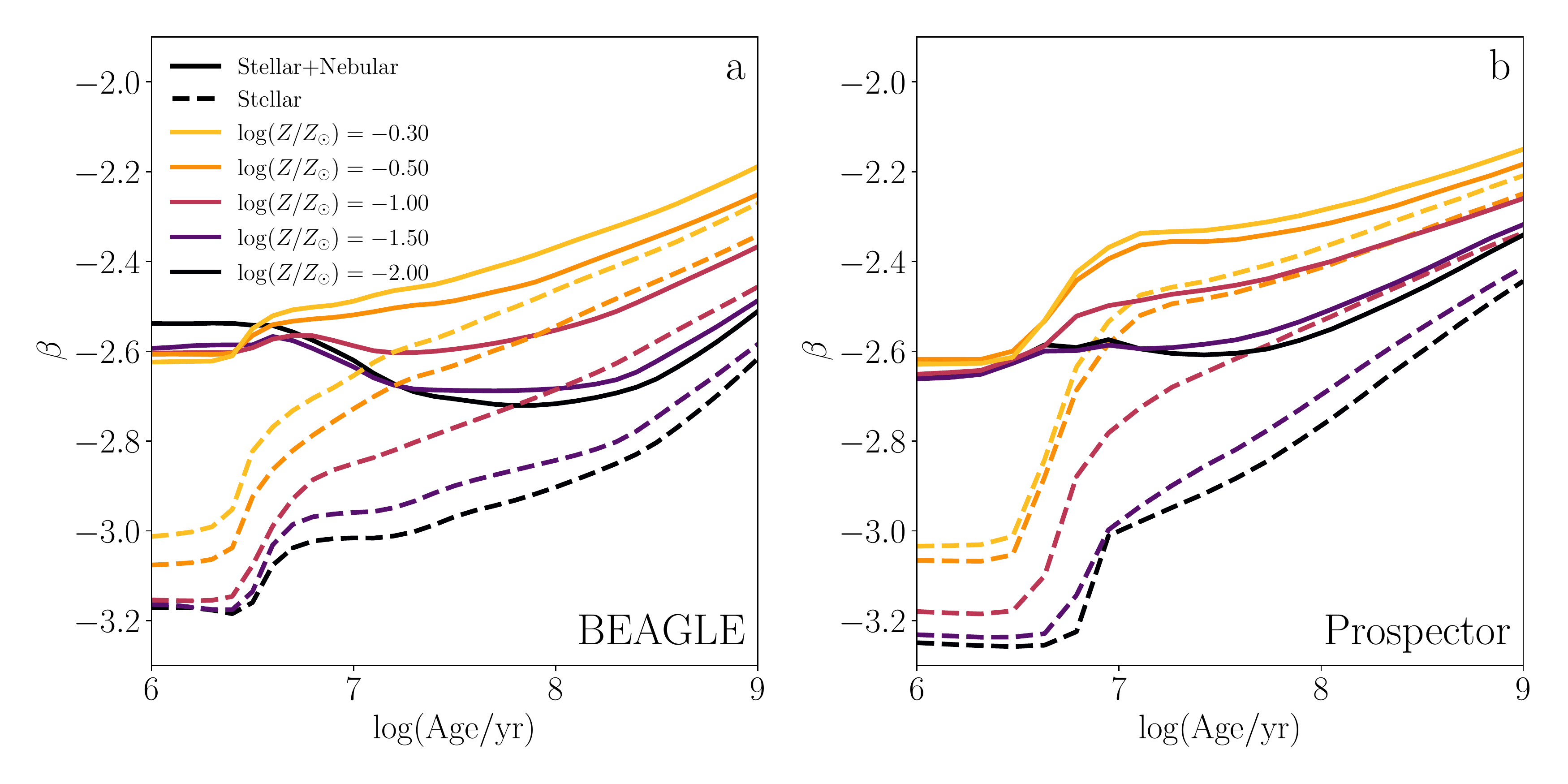}
    \caption{(\textit{a}): Rest-UV continuum slope, $\beta$, as a function of age and metallicity measured from models constructed using BEAGLE.  These models assume a constant star-formation history, solar abundance patterns, $\log(U)=-2.5$, and contain no dust attenuation.  Shown here are results from models of stellar light only (dashed lines), and those with self-consistently computed nebular continuum emission (solid lines).  As described in Section~\ref{sec:uvslopes}, the addition of nebular emission reddens the UV continuum by $\delta \beta\sim0.6$ at the youngest ages and lowest metallicities, and has a decreasing effect toward older galaxies and higher metallicities.  (\textit{b}): Same as panel (a) except the UV slopes are calculated using models produced using Prospector.   We note that while we display UV slopes of models with ages up to 1Gyr for reference, this extends beyond the age of the universe at the redshifts considered in this study.}
    \label{fig:BEAGLE-uvslopes}
\end{figure*}

To guide our investigations in the following section, we also quantify how UV slope varies with stellar population parameters for the fiducial BEAGLE models we are using. This has been done previously for other stellar population models (e.g., \citealt{Bouwens2010, Dunlop2013, Zackrisson2013}), but we seek to test whether there are significant differences in the intrinsic UV slopes for the updated models we are considering in this paper. Figure~\ref{fig:BEAGLE-uvslopes} shows how the power-law slope, $\beta$, varies as a function of age and metallicity, calculated as the power-law slope over the range $1280-2580$\AA\ as in \citet{Calzetti1994}.  This figure includes UV slopes calculated using BEAGLE models of the stellar emission only, as well as models that include the contribution from the nebular continuum.  The physics driving variations in the UV slope with age and metallicity has been described elsewhere (e.g. \citealt{Reddy2012b,Topping2015}).  While we impose a \citet{Chabrier2003} IMF in this study, we note that different assumptions, such as a top-heavy IMF, can impact the relative abundances of the most massive stars, affecting the stellar continuum shape. Results from \citet{Wilkins2012} suggest that imposing a top-heavy IMF with a high-mass slope of $\alpha=-1.5$ could result in a steepening of the UV slope of $\Delta\beta\sim0.2$ compared to when a high-mass slope of $\alpha=-2.35$ is assumed.
\citet{Jerabkova2017} further explored how UV slopes vary under different IMF assumptions as a function of age. They demonstrated that at all but the youngest ages (i.e., $\lesssim5$Myr), the choice of IMF played a minor role in the UV slope, but observed a similar steepening to that found by \citet{Wilkins2012}, amounting to a $\Delta\beta\sim0.2$ at the youngest ages.
These results suggest that a varying IMF plays an important role in setting the UV slope, however the leakage of ionizing photon is still needed to achieve the bluest slopes in our sample.

The BEAGLE models asymptote to $\beta = -3.2$ at the youngest ages ($< 3$ Myr) and lowest metallicities ($\log(Z/Z_{\odot})=-2$) considered. We note similarly blue UV slopes are seen in the Prospector models \citep{Johnson2021} with the default FSPS stellar templates \citep{Conroy2009, Conroy2010} which implement the MIST stellar isochrones \citep{Choi2016} (see right panel of Figure ~\ref{fig:BEAGLE-uvslopes}). Here we have adopted the same model set-up as in our fiducial BEAGLE models.
These results thus indicate that BEAGLE and Prospector have no problem reproducing UV slopes as blue as $\beta \simeq -3$, provided the metallicities are low, ages are young, and stellar continuum dominates the light. 
In addition to age and metallicity, stellar multiplicity can be an additional factor regulating the shape of the UV continuum.  We explore the possibility that the introduction of binary stars could yield significantly bluer UV slopes using stellar templates from BPASS \citep{Eldridge2017, Stanway2018}.  However, the BPASS models reach a minimum $\beta$ of $-3.15$, similar to that found using BEAGLE and Prospector, indicating our fiducial models capture the range of possible UV slopes under a variety of assumptions.
The solid lines in Figure ~\ref{fig:BEAGLE-uvslopes} demonstrate the reddening effect of nebular continuum in the BEAGLE and Prospector models. The impact of nebular continuum is strongest at young ages and low metallicities (see Figure ~\ref{fig:nebularfrac}), causing the intrinsic UV slope (i.e., before dust attenuation) 
to approach $\beta\simeq -2.6$ at the youngest ages.  We note that once nebular continuum has been added, the intrinsic UV slopes are not strongly dependent on 
age (for a constant SFH), with older models ($\simeq 100$ Myr) fully capable 
of reproducing very blue UV slopes, provided minimal attenuation is present.

\begin{figure*}
    \centering
    \includegraphics[width=1.0\linewidth]{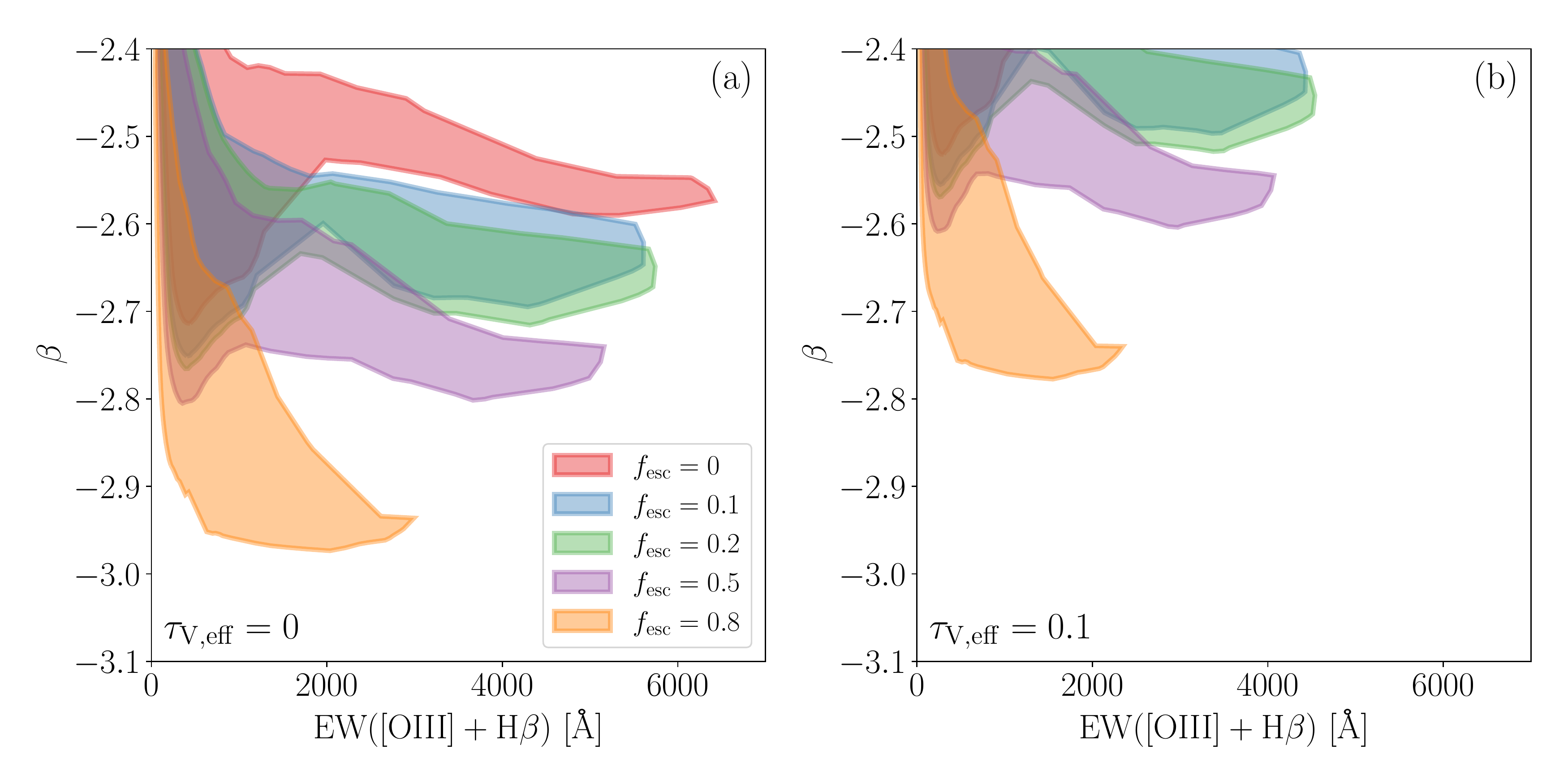}
    \caption{{\it a:} Regions of the $\beta$-EW parameter space probed by SED models with varying amounts of \fesc.  These regions were derived by constructing grids of SED models spanning a broad range of ages and metallicities  (i.e., $\log(\rm Age/yr)=6\rm\ to\ 9)$; $\log(Z/Z_{\odot})=-2\rm\ to\ 0.3$). Models with an \fesc=1 will trend toward bluer $\beta$, down to a minimum of $-3.2$ (see Figure~\ref{fig:BEAGLE-uvslopes}), but do not produce any line emission (i.e., EW). As such, they are not displayed on this figure. {\it b:} Same as left panel, except calculated from BEAGLE models with a modest attenuation from dust, corresponding to $\tau_{\rm V,eff}=0.1$. The addition of dust in these models results in reddened UV slopes and decreased EWs. \vspace{10pt}}
    \label{fig:beta-ew}
\end{figure*}

It is clear that according to fiducial models within BEAGLE and Prospector, selecting galaxies with $\beta < -3$ will pick out a population with moderately low stellar metallicity and significant leakage of ionizing photons from the HII regions and this threshold motivates our search in \S3.  To self-consistently model sources we find, we include nebular emission in conditions with non-zero \fesc{}, adopting models with varying amounts of ionizing radiation escape calculated in the density-bounded regime from \citet{Plat2019}. 
These models span a large range of \fesc{}, up to extreme values of 50\%-80\%, consistent with constraints of the most extreme leakers observed at lower redshift \citep[e.g.,][]{Vanzella2018}.
To test how imposing varying amounts of \fesc\ on the model SEDs impacts the resulting UV slopes 
and nebular emission line strengths, we construct grids of models at fixed \fesc\ spanning the full range of ages (1 Myr--1 Gyr), and metallicities ($\log(Z/Z_{\odot})=-2$ to $0.3$) available, and assuming no dust.  

The resulting ranges of $\beta$ and [OIII]+H$\beta$ EW found throughout these variable \fesc grids are shown in Figure~\ref{fig:beta-ew}.  As expected, increasing \fesc\ allows models to be constructed that yield bluer UV slopes.  Additionally, larger values of \fesc\ restrict the maximum EWs that can be observed, in line with the results from previous studies \citep[e.g.,][]{Zackrisson2013, Plat2019}. However we note that strong emission lines are still present in some cases with significant leakage of ionizing photons. For example, we find [OIII]+H$\beta$ EWs of over 1000~\AA\ for young ages (1-3 Myr) and moderate metallicities (0.2-0.5 Z$_\odot$) with significant leakage (\fesc=0.8). This follows naturally from the fact that emission lines for these young sources are extremely strong in BEAGLE models, and hence the lines remain moderately strong even if the flux has been reduced by 
the escape of ionizing photons. It is only at lower gas-phase metallicities and slightly older ages where 
the lines become considerably weaker in the high \fesc models. We thus expect very blue UV slopes to flag sources with young stellar ages, large values of \fesc, and the strength of the nebular lines then dictates whether the metallicity is very low (weak lines) or moderate (stronger lines).

The model predictions outlined in Figure~\ref{fig:beta-ew} (b) demonstrate the importance of assuming no dust attenuation for this selection.  For example, even the presence of dust corresponding to a $\tau_{\rm V,eff}=0.1$ reddens the UV continuum yielding a UV slope that is indistinguishable from an un-attenuated SED with \fesc$=0$ for all but the most extreme values of \fesc{}. As such, focusing our selection toward the bluest UV slopes provides the best chance of identifying objects with large \fesc{}. Additionally, while the increased presence of dust can lead to weaker emission lines, the amount of attenuation required will require a significant reddening of the UV continuum to values outside our selection. Variation to the strength of these emission lines can also occur due to changes to the IMF \citep[e.g.,][]{Zackrisson2013}.
However, assuming an IMF resulting in bluer UV slopes would likely strengthen the emission lines, making such objects less attractive as high \fesc{} candidates based on our selection.

Based off what we have reviewed in this section, our selection in the NIRCam 
imaging of the EGS field will focus on galaxies with $\beta\simeq -3$, taking care to ensure that the slope is blue in the FUV and NUV. We will then look to the flux excesses in the rest-optical to probe the rest-optical line strengths, but we will not rule sources out if they have moderately strong 
line emission.

\begin{figure*}
    \centering
    \includegraphics[width=1.0\linewidth]{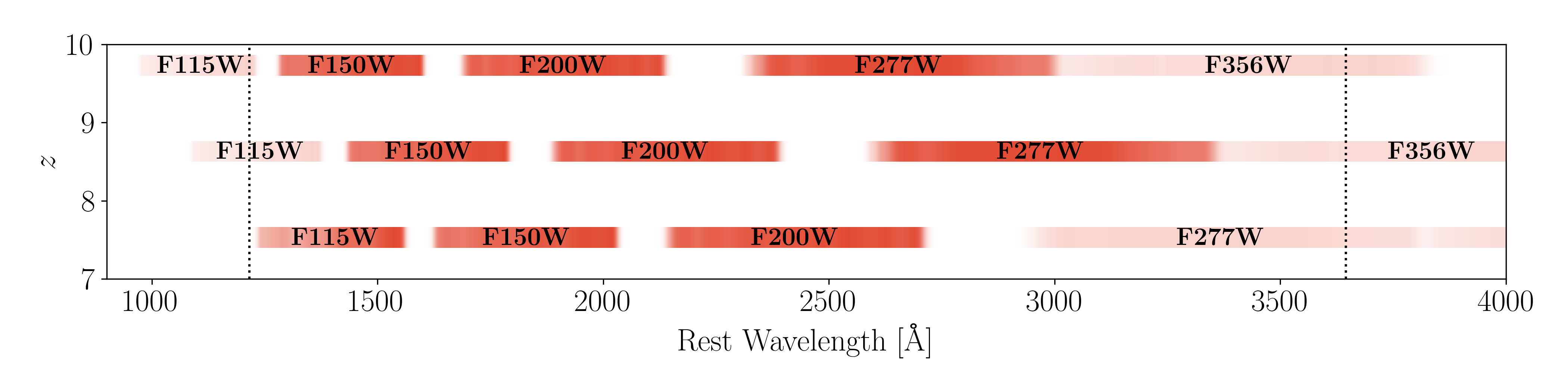}
    \caption{Rest-frame coverage of available NIRCam filters throughout the rest-UV continuum.  Bands are shown at three different redshifts spanning the different photometric selections described in Section~\ref{sec:data}.  The position of Ly$\alpha$ and the Balmer break are displayed as the vertical lines.  Filters that extend past these boundaries, and thus are less sensitive to the UV continuum, are displayed at a reduced opacity.}
    \label{fig:betafilts}
\end{figure*}

\begin{figure*}
    \centering
    \includegraphics[width=1.0\linewidth]{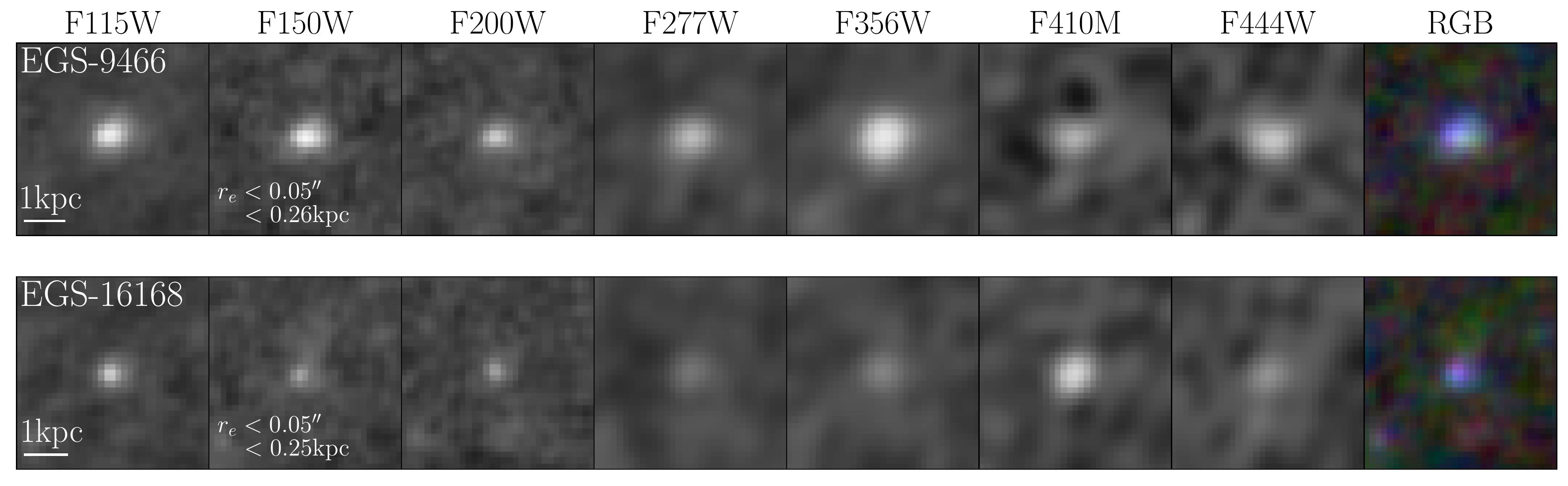}
    \caption{Image cutouts in all NIRCam filters and combined color image of the two extremely blue objects in the sample. Postage stamps are all $1^{\prime \prime}$ on a side centered at the location of the object. All of the objects present clear detections in multiple filters indicating they are not artifacts. }
    \label{fig:postage}
\end{figure*}

%
%
\section{Observations and Analysis}
\label{sec:3}

\subsection{NIRCam Imaging and Photometry of the EGS Field}
\label{sec:data}
We utilized imaging obtained as part of the first data release from the Cosmic Evolution Early Release Science \citep[CEERS][]{Finkelstein2017} survey\footnote{\url{https://ceers.github.io/}}. These data include deep \jwst/NIRCam imaging over the Extended Goth Strip (EGS) observed during June 2022 in three short-wavelength (SW) bands (F115W, F150W, and F200W), in addition to four long-wavelength (LW) bands comprising three wide and one medium band (F277W, F356W, F410M, and F444W). 
A detailed description of the reduction procedure is provided in  \citet{Whitler2022b} and \citet{Endsley2022e}, however we give a brief summary here.
Calibrated co-added mosaics for each band were produced by first processing individual detector exposures through the JWST pipeline\footnote{\url{https://jwst-pipeline.readthedocs.io/en/latest/index.html}} after implementing $1/f$ noise subtraction and background subtraction.
We adopt photometric zero points derived by Gabe Brammer\footnote{https://github.com/gbrammer/grizli/pull/107} calculated using observations of a flux calibration standard in NIRCam for a subset of filters in module B, in addition to NIRISS and NIRCam observations of the LMC astrometric calibration field for the remaining filters and module A. These zero points have been shown to produce reasonably consistent color-magnitude diagrams from the JWST Resolved Stellar Populations ERS program \citep{Boyer2022}.
The co-added images of individual filter, detector, and pointing combinations were then matched to the \textit{Gaia} astrometric frame by running \textsc{tweakreg} using the \textit{Gaia}-aligned \hst{}/WFC3 F160W CHArGE mosaic (see below) as a reference image (see \citet{Chen2022} for more details). The resulting alignment achieves an rms offset of $\sim6-15$ mas compared to the \textit{Gaia}-aligned \hst{} imaging. The calibrated and aligned NIRCam images for each detector and pointing were then co-added into final mosaics for each band, and resampled onto a common pixel grid with 0\secpoint03/pixel spacing. 

We additionally utilized deep observations from \hst\ collated as part of the Complete Hubble Archive for Galaxy Evolution (CHArGE). Specifically, this additional data comprises \hst/ACS F435W, F606W, and F814W obtained as part of the All-Wavelength Extended Goth Strip International Survey \citep[AEGIS;][]{Davis2007}, Cosmic Assembly Near-infrared Deep Extragalactic Legacy Survey \citep[CANDELS;][]{Grogin2011, Koekemoer2011}, and UVCANDELS (PI: Teplitz \footnote{\url{https://archive.stsci.edu/hlsp/uvcandels}}) surveys.  Imaging data from these programs were combined using the software package \textsc{grizli} \citep{Brammer2022}, and then subsequently registered to the \textit{Gaia} coordinate system and sampled at a consistent pixel scale of 0\secpoint04 /pixel.  Finally, to provide consistent photometric measurements across all filters, the \jwst/NIRCam LW bands were PSF-matched that of WFC3/F160W mosaics (also from CHArGE; 0\secpoint08 /pixel), while the \jwst/NIRCam SW and \hst/ACS images were adjusted so that their PSF-matched that of the ACS/F814W image. Details regarding this PSF-matching procedure largely follow that of \citet{Endsley2021} and are discussed in more detail in  \citet{Endsley2022e} .

Photometry on this combined \hst/ACS and \jwst/NIRCam dataset were calculated using elliptical apertures constructed to capture the full galaxy emission in each of the PSF-homogenized SW and LW bands.  Errors on the photometry were estimated by placing many iterations of the same sized aperture among blank regions of the sky surrounding each object.  Corrections for contamination from nearby bright objects were made by fitting models to the flux distribution of the contaminant, and subtracting off the excess emission.  Further details regarding the photometric measurements are provided in  \citet{Whitler2022b} and \citet{Endsley2022e}.

\begin{figure}[t]
    \centering
    \includegraphics[width=1.0\linewidth]{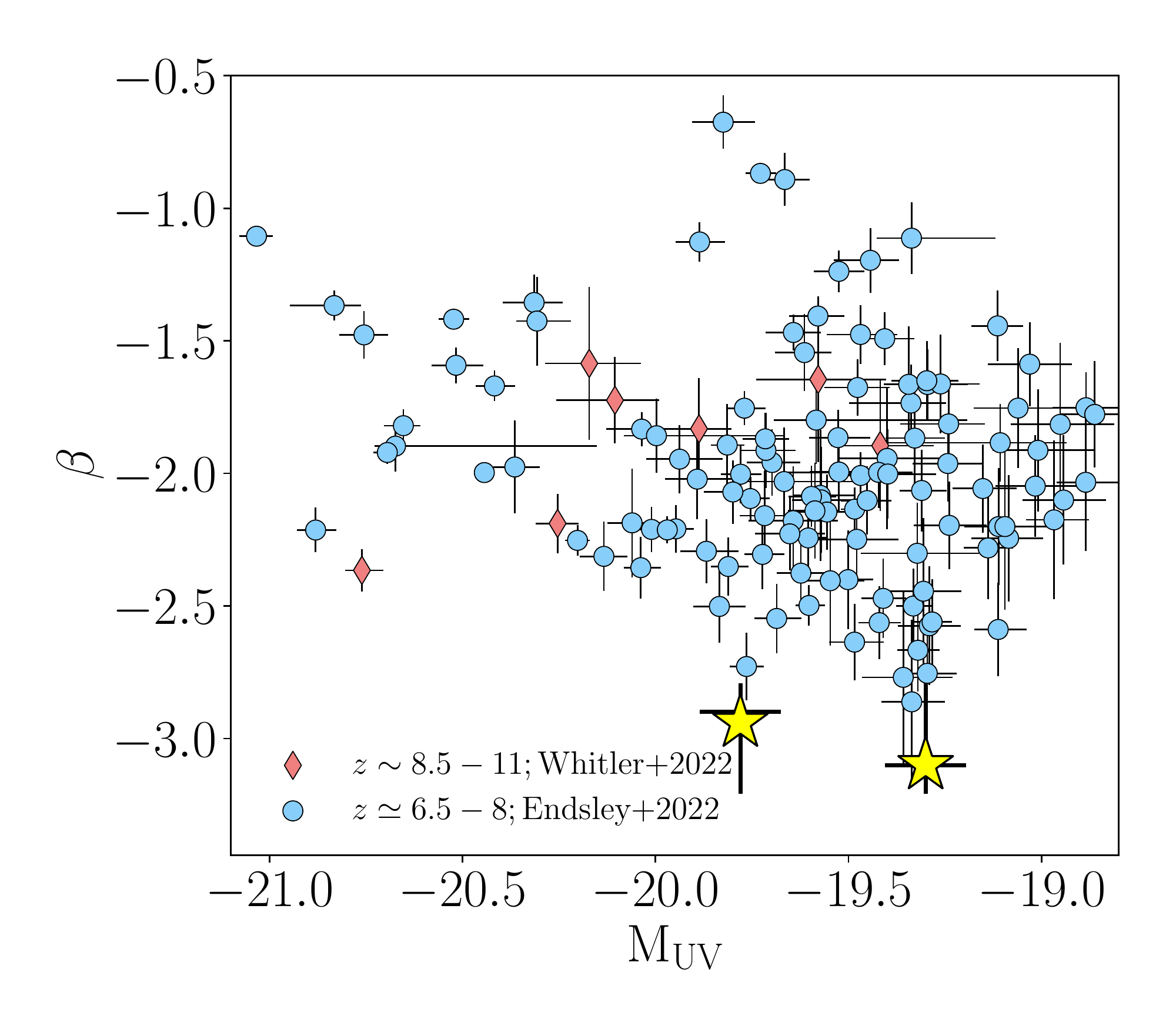}
    \caption{UV slopes as a function of $\rm M_{\rm UV}$ for the full sample of $z\sim7$ (blue circles; \citealt{Endsley2022e}) and $z\sim8.5-11$ (red diamonds; \citealt{Whitler2022b}) galaxies.  The two objects identified with extremely blue UV slopes, and described in Tables~\ref{tab:objects} and \ref{tab:properties} are shown as the yellow stars. }
    \label{fig:fesc}
\end{figure}

\subsection{Selection of $z>7$ Galaxies with Very Blue UV Slopes}
The  deep observations provided by \jwst/NIRCam allow robust selections within several redshift windows at $z>7$. Our group has developed selections of galaxies at redshifts spanning $z\sim6.5-11$, comprising two photometric dropout selections at $z\sim6.5-8$ and $z\sim8.5-11$.  A detailed discussion of these selections are provided in  \citet{Whitler2022b} and \citet{Endsley2022e}, for the lower and higher redshift windows, respectively, and we provide a brief description here.
Initial selection of $z\sim6.5-8$ galaxies  \citep{Endsley2022e} was obtained by objects that satisfy the following conditions:
\begin{eqnarray*}
    (\textrm{F200W}<28)\ \wedge\ (\textrm{F606W}-\textrm{F115W}>1.5)\ \wedge \\ 
    (\textrm{F814W}-\textrm{F115W}>1.5) \wedge (\textrm{F115W}-\textrm{F200W}<0.5) \wedge \\
    (\textrm{F814W}-\textrm{F115W}>\textrm{F115W}-\textrm{F200W}+1.5).
\end{eqnarray*}
For the purposes of these color cuts, the flux in the dropout bands (i.e. F606W and F814W) are set to their 1$\sigma$ upper limit in cases of non-detections (S/N$<$1).
We additionally require non-detections ($\rm S/N<2$) in ACS/F435W, F606W, and F814W, and optical $\chi^2$ \citep{Bouwens2015} of $\chi^2_{\rm opt} < 5$ calculated using the three ACS bands. A $\rm S/N>3$ detection is required in at least two \jwst/LW bands, with a constraint of a $\rm S/N>5$ detection in F200W.  

A selection of $z\sim9-11$ galaxies  \citep{Whitler2022b} is obtained through the following photometric conditions ensuring placement of the Ly$\alpha$ break:
\begin{eqnarray*}
    (\text{F115W} - \text{F150W} > 1.0) \wedge 
    (\text{F150W} - \text{F277W} < 0.4) \wedge \\
    (\text{F115W} - \text{F150W} > 0.8 \times \text{F150W} - \text{F277W} + 1.5),
\end{eqnarray*}
with an added requirement of non-detection ($\rm S/N<2$) in the ACS bands, and a detection at the $> 7\sigma$ level in F200W, and $> 3\sigma$ for all LW bands. Similar to the approach above, the flux in the F115W dropout band is set to its 1$\sigma$ upper limit in the case of non-detection.

This high-redshift sample is supplemented in \citet{Whitler2022b} by a selection at slightly lower redshift ($z\sim8.5-9$), where the observed-frame Ly$\alpha$ break yields a partial decrease in the emission within F115W, and may not suppress the entirety of the flux in this band. Specifically, galaxies in this redshift window are identified by the following criteria:
\begin{eqnarray*}
    (\text{F115W} - \text{F150W} > 0.6) \wedge
    (\text{F150W} - \text{F277W} < 0.4) \wedge \\
    (\text{F115W} - \text{F150W} > 1.5 \times \text{F150W} - \text{F277W} + 0.6).
\end{eqnarray*}
As before, we additionally require a $> 7\sigma$ detection in F200W, and a $> 3\sigma$ detection in all LW bands, with the further requirement of a $<28$ mag in F200W, and non-detection  ($<$2$\sigma$) in the ACS bands. 
We characterize photometric redshift probability distributions using BEAGLE, removing those sources with large probabilities ($>$30\%) of being at $z<8.2$.

\begin{figure*}
    \centering
    \includegraphics[width=1.0\linewidth]{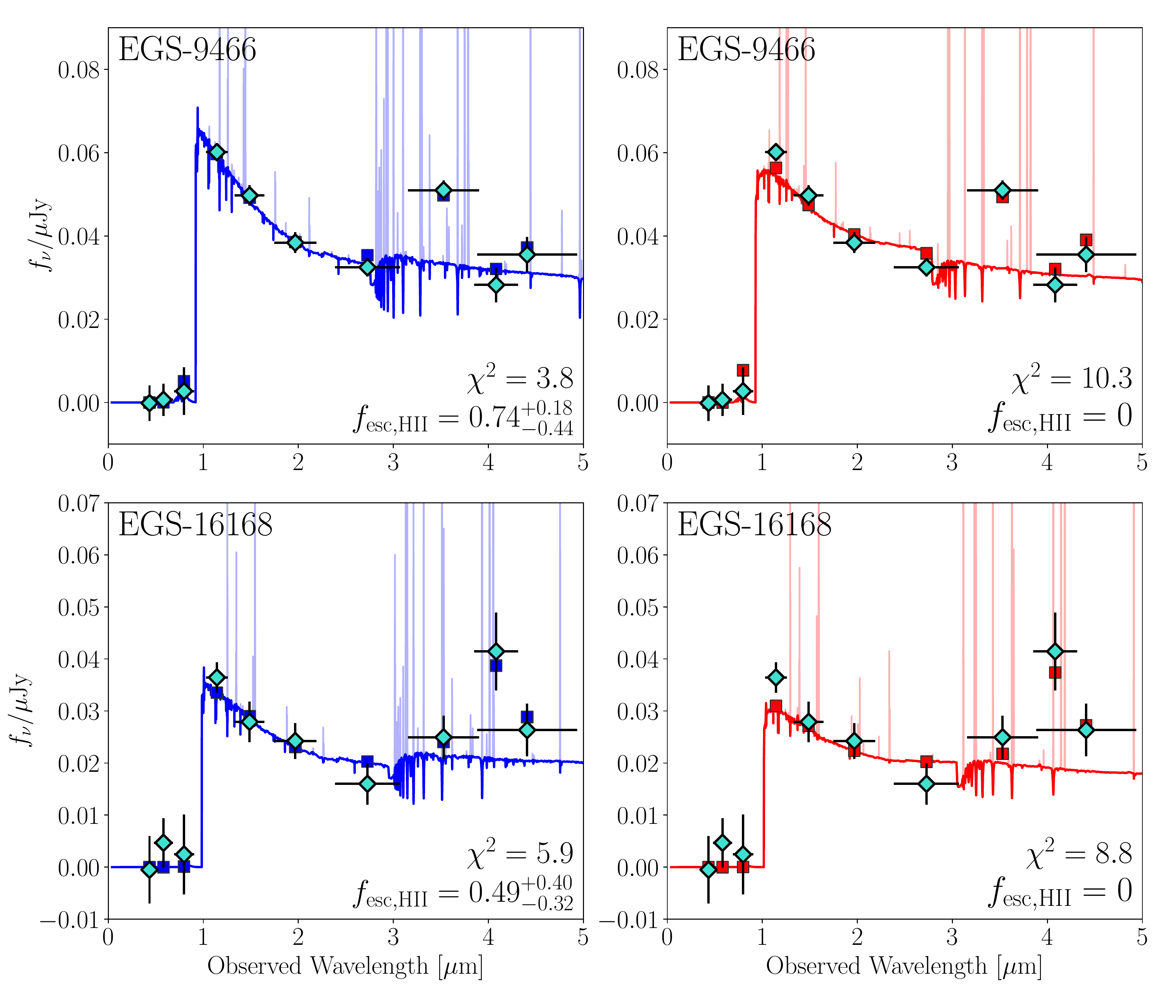}
    \caption{Best-fit SEDs for the two extremely blue objects in our sample. The best-fit SED calculated from BEAGLE using models that allow for the escape of ionizing radiation is displayed in the left panel, while the model with fixed \fesc$=0$ is given on the right.  In both panels, the observed photometry are displayed as the cyan diamonds, while the predicted photometry from the models are displayed as colored squares. The minimum $\chi^2$ found for each set of models, as well as the best-fit \fesc\ are given in the lower right of each panel.  }
    \label{fig:seds}
\end{figure*}

We derive properties for the 123 galaxies in our $z\simeq 7-11$ sample using BEAGLE \citep{Chevallard2016}. We broadly follow the approach described in \S2.  For galaxies selected as part of the $z\sim7$ selection, we assume uniform priors in log space on both metallicity and ionization parameter, sampled over the full range allowed by the models (i.e., $\log(Z/Z_{\odot})\in[-2.2, 0.3]$; $\log(U)\in[-4,-1]$).  Our models assume an SMC dust law \citep{Pei1992}, which has been shown to reproduce the observed properties of high-redshift galaxies \citep[e.g.,][]{Bouwens2016, Reddy2018}.  All models are constructed using a constant star-formation history (SFH), and we impose a log-uniform prior on the assumed age ranging from 1Myr to the age of the universe at the given redshift. We define these as our fiducial models and will consider the influence of varying the escape fraction using density-bounded models in \S4. A detailed discussion comprising the full set of properties derived for the $z\sim7$ and $z\gtrsim8.5$ samples is provided in \citet{Endsley2022e} and \citet{Whitler2022b}, respectively.

To measure $\beta$ from the observed photometry, we utilize available NIRCam filters that are sensitive to the rest-frame UV continuum.  Crucially, we avoid filters that may include emission from Ly$\alpha$ or may be biased by including wavelengths in the Ly$\alpha$ forest. We additionally exclude filters that extend redward of the Balmer break in the rest frame, as they may be biased due to an older stellar population or other effects.
Figure~\ref{fig:betafilts} illustrates the rest-frame wavelengths probed by the available filters, in addition to examples of the selection broadband filters used at several redshifts. 
UV slopes are then calculated for each object by a power-law fit to the observed photometry that is confined to wavelengths bounded by Ly$\alpha$ and the Balmer break.
Uncertainties on $\beta$ are calculated using Monte-Carlo simulations, where both the observed photometry is perturbed by their corresponding errors, and the photometric redshifts are sampled from the derived posterior probability distribution. 
With each iteration of these simulations, the set of filters considered to probe the UV continuum are re-established due to the changing redshift.

Figure~\ref{fig:fesc} provides derived UV slopes for our sample as a function of their absolute UV magnitude. 
On average, we find that the galaxies in our sample have blue UV slopes, represented by a median value of $-2.0$ for the $z\simeq 7$ sample and $ -1.9$ for the $z\simeq 8.5-11$ sample, with the reddest galaxies reaching values of $\beta=-0.7$.
 These results are consistent with other measurements of $z\gtrsim8$ galaxies derived from NIRCam imaging, which cover a similar range of UV luminosity \citep[e.g.,][]{Cullen2022}.
We note that these are averages of our measured data and should not be used 
as averages of the population. Such analysis requires comprehensive simulations to account for systematics of the selection function \citep[e.g.,][]{Dunlop2012, Rogers2013, Bouwens2014} which are beyond the scope of the present 
work. 
Two galaxies within these high-redshift selections stand out as having UV slopes significantly bluer than the majority of the sample. These two systems have derived photometric redshifts of $z_{\rm phot}=6.7$ and $z_{\rm phot}=7.2$, and show UV slopes of $ \beta=-2.9^{+0.1}_{-0.3}$, and $ \beta=-3.1^{+0.3}_{-0.1}$, respectively (see Table~\ref{tab:objects}). Each of these systems matches the selection motivated in \S2 and are good candidates for LyC leakers with reasonably low metallicities. For both galaxies, solutions at low redshift ($z<6$) comprise a minimal fraction (<5\%) of the total redshift probability distribution. We note that galaxies with UV slopes as blue as $\beta\sim-3$ have been found in other samples of high-redshift galaxies selected from NIRCam imaging \citep[e.g.,][]{Atek2022, Castellano2022, Furtak2022}.

The observed SEDs are presented in Figure~\ref{fig:seds}. In each object, the blue UV slope is clearly visible across at least three NIRCam filters, extending across both the FUV and NUV. This helps 
reduce the likelihood that the UV slopes arise due to random scatter, while also satisfying expectations for minimal contribution from nebular continuum, as would be expected 
if there is significant ionizing photon leakage (see \S2). 
The SEDs of these reveal weak-to-moderate photometric excesses in F356W where strong nebular emission lines (i.e., [OIII]+H$\beta$) contribute to the flux. 
The inferred strength of the rest-optical emission lines, in combination with the exceedingly blue UV slopes are preferentially fit with SED models that allow for the escape of ionizing radiation. 
The models describing these galaxies, as well as implications for \fesc and metallicity, are described in detail in the next section.

While these extremely blue sources possess moderately weak [OIII]+H$\beta$ EWs, a population of very young (e.g., $<10$Myr) galaxies with even weaker line emission has emerged at $z\sim6.5-8$ \citep{Endsley2022e}.
While no objects in this population, which comprise $\sim20\%$ of the $z\sim6.5-8$ selection, exhibit extremely blue UV slopes, the weak line emission is a characteristic of LyC photon escape. 
While non-zero \fesc{} models are not strongly preferred by BEAGLE, they do still provide an accurate description of this photometry.
Specifically, when \fesc{} is fixed to zero, the weak lines are best reproduced by extremely low metallicities ($\sim1\%$).  In this regime the SED fits arrive at a typical minimum $\chi^2$ of 4.5.
When non-zero \fesc{} is assumed, there is more flexibility in the inferred properties while still being able to fit the lines.
At high values of \fesc, we still achieve a typical minimum $\chi^2$ of 5.0, indicating fits with similar fidelity to the \fesc{}$=0$ models.  
The BEAGLE fits indicate the presence of dust, corresponding to a $\tau_{\rm V,eff}=0.13$, preventing UV slope from reaching values of $\sim -3$.
So while these objects do not possess the {\it combined} extremely blue UV slopes and low emission line strengths to confidently suggest high levels of \fesc, such properties may still be present in these systems. 
A complete discussion of the potential characteristics of this population is provided in \citet{Endsley2022e}.

\begin{table}
\begin{center}
\begin{tabular}{cccccc}
\toprule
Object ID & RA & Dec & $z_{\rm phot}$ & F200W &$\beta$  \\
\midrule

 EGS-9466 &   $214.77676$ & $+52.84293$   & $6.7^{+0.1}_{-0.1}$ & 27.9 & $-2.9^{+0.1}_{-0.3}$  \\
 \\
  EGS-16168 & 214.79312 & $+$52.87012  & $7.2^{+0.1}_{-0.2}$  & 27.4 & $-3.1^{+0.3}_{-0.1}$  \\
  
 \bottomrule
 \end{tabular}
 \end{center}
\caption{Properties of the extremely blue sample. Photometric redshifts are derived using BEAGLE models.  As described in \S3.2, UV slopes are calculated by fitting a power-law slope to the photometry residing within the rest-UV.}
\label{tab:objects}
\end{table}

\begin{table*}
\begin{center}
\begin{tabular}{ccccccccccc}
\toprule
Object ID & F435W & F606W & F814W & F115W & F150W & F200W & F277W & F356W & F410M & F444W  \\
\midrule

 EGS-9466 & $-0.1 \pm 4.0$ & $0.6 \pm 3.6$  & $2.7 \pm 5.4$  & $60.1 \pm 1.9$ & $49.7 \pm 2.2$ & $38.4 \pm 2.2$ & $32.5 \pm 1.9$ & $51.0 \pm 2.1$ & $28.2 \pm 3.9$ & $35.5 \pm 4.0$ \\
 \\
  EGS-16168 & $-0.5 \pm 6.2$ & $4.7 \pm 4.5$ & $2.4 \pm 7.5$ & $36.4 \pm 2.7$ & $27.9 \pm 3.7$ & $24.2 \pm 3.2$ & $16.0 \pm 3.8$ & $24.9 \pm 3.9$ & $41.4 \pm 7.3$ & $26.3 \pm 4.8$  

 \\

 \bottomrule
 \end{tabular}
 \end{center}
\caption{ Flux measurements and errors for the two galaxies identified with very blue UV-slopes in nJy. The derivation of these fluxes are described in \S~\ref{sec:data}.}
\label{tab:flux}
\end{table*}

\section{The Physical Properties of $z>7$ Galaxies with Extremely Blue UV Slopes}

In the previous section, we identified two objects in $z\simeq 7-11$ galaxy selections that have extremely blue ($\beta\simeq -3$) UV continuum slopes. In this section we explore the properties of these systems in more detail.  We first consider properties 
derived from our fiducial (ionization-bounded) BEAGLE 
models. We follow the same model setup as described in \S3 for our initial photometric redshift runs to characterize the properties of our galaxies. 
Namely, we apply a uniform prior in redshift restricting the allowed values to $z=[5,12]$, and impose a log-uniform prior on metallicity in the range $\log(Z/Z_{\odot})\in[-2.2, 0.3]$, and allow the ionization parameter to vary within $\log(U)\in[-4,-1]$ with a uniform prior.
We also fit each object using ionization-bounded models that approximate the effects of alpha-enhancement, such that the metallicity of the stars (set by iron and iron-peak elements) is lower than that of the gas (set by oxygen and other alpha elements). And we consider the density-bounded models with varying \fesc described in \S2 (see \citealt{Plat2019}).
An overview of the derived properties of these two galaxies is provided in Table~\ref{tab:properties}, and discussed below.

In Figure~\ref{fig:seds}, we display the best-fit SED using the models that allow for the escape of ionizing radiation from HII regions, as well as the results when we fix \fesc$=0$.  
In both objects, the models with varying \fesc\ provide much better fits to the observed photometry. 
This agreement is apparent both visually and based on the $\chi^2$, such that for EGS-9466 and EGS-16168 the varying \fesc\ models provide a $\chi^2=3.8$ and $\chi^2=5.9$, while the \fesc$=0$ models only achieve $\chi^2=10.3$ and $\chi^2=8.8$, respectively.
As pointed out in the previous sections, simultaneously reproducing the shape of the UV continuum and relative weakness of the emission requires a reduction in the amount of nebular emission contributing to the flux.
The density-bounded models are able to reproduce both the shape of the UV continuum and inferred strength of the emission lines simultaneously, which is not possible under the constraint of \fesc$=0$. In the fiducial (ionization-bounded) models the UV slopes 
are clearly very poorly matched owing to the contribution of nebular continuum.

The increased ability to explain the observed photometry with these updated models allows us to explore the stellar populations within these extreme objects at high redshift.  
In addition to allowing \fesc\ to vary within the model galaxy spectra, other exotic populations, such as Pop III stars, have been introduced as potential solutions to poorly-fit extremely blue SEDs.  
We find that the observed photometry of these galaxies is well-fit through only a high \fesc{}, and such exotic populations are not necessary. 
We additionally fit the SEDs of these objects using models that are $\alpha$-enhanced (i.e., $Z_{\star} < Z_{\rm ISM}$), corresponding to conditions observed in high-redshift galaxies \citep[e.g.,][]{Steidel2016, Sanders2020, Topping2020a, Strom2022}. Within these models, the stellar and gas-phase metallicities are decoupled, thus enabling a hard ionizing spectrum (from low-metallicity stars) to irradiate ISM with moderate metallicities, thereby yielding stronger lines. Specifically, we combined \citet{cb2003} stellar templates with nebular emission calculated with \citet{Ferland2017}, and tested the case where the stellar metallicity is fixed to $0.2Z_{\rm ISM}$ (corresponding to the theoretical limit of Type II SN enrichment), and and also where the stellar metallicity is fixed to $Z_*=10^{-4}$, and $Z_{\rm ISM}$ is allowed to vary. However, models allowing for alpha-enhancement, but without the escape of ionizing radiation, are not able to explain the observed photometry.
The high level of inferred \fesc\ implies that the UV continuum is increasingly sensitive to continuum emission from the most massive stars. This sensitivity in the absence of nebular contamination provides an improved view into the properties of the massive star population.
 The BEAGLE SED models overwhelmingly prefer solutions with low stellar metallicities, such that we find a value of $ \log(Z/Z_{\odot})=-1.6$ and $\log(Z/Z_{\odot})=-1.4$ for EGS-9466 and EGS-16168, respectively.  
As suggested in \S2, these low metallicities are required to reproduce the observed color excesses in the long wavelength filters, which correspond to [OIII]+H$\beta$ EWs of 440\AA{} and 410\AA{} for EGS-9466, and EGS-16168, respectively.
High-mass stars at such extremely low metallicities would be intense sources of ionizing radiation that, in combination with large values of \fesc, would make these objects highly efficient ionizing agents at this epoch.

\begin{figure*}
    \centering
    \includegraphics[width=1.0\linewidth]{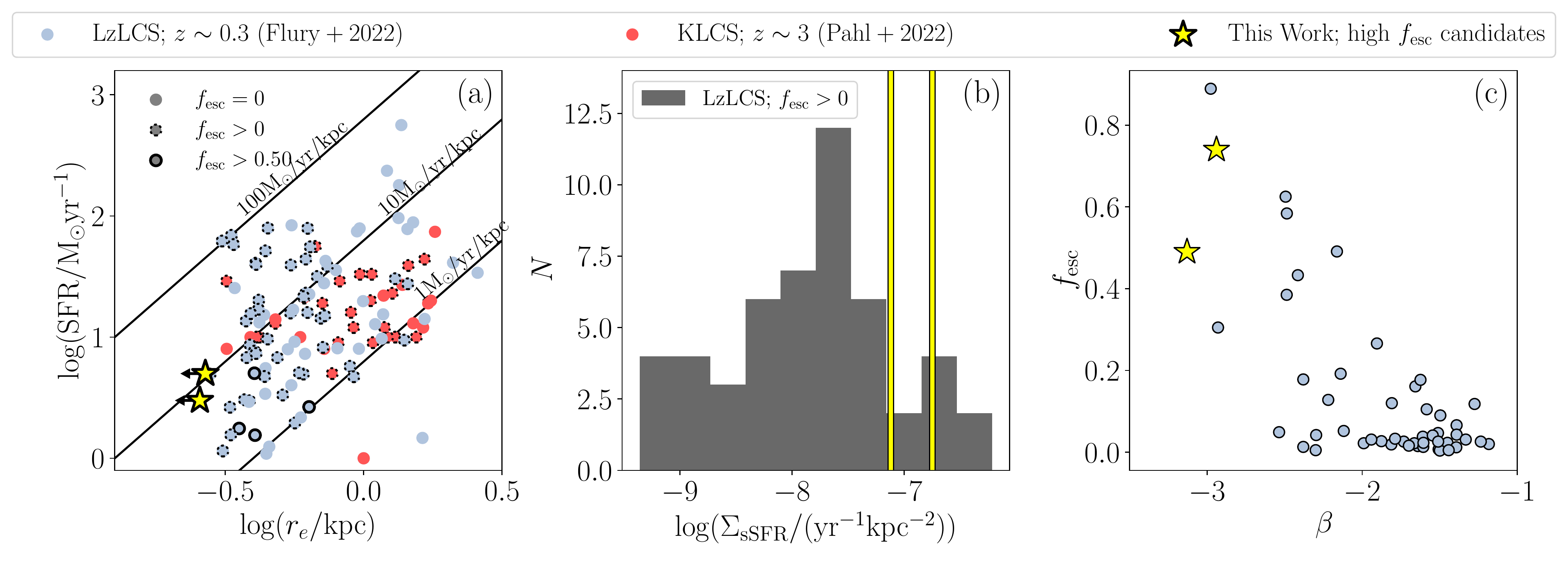}
    \caption{Comparison of our two high-\fesc candidates with low-redshift samples.  {\it a:} SFR vs effective radius for the two objects in our sample (yellow stars), the Keck Lyman Continuum Survey sample \citep[KLCS, red circles; ][]{Steidel2018, Pahl2022}, and the Low-$z$ Lyman Continuum Survey \citep[LzLCS, blue circles;][]{Flury2022a}. Lines of constant SFR surface density are shown for reference. {\it b:} Distribution of sSFR surface density ($\Sigma_{\rm sSFR}$) for the LzLCS sample. The values for the two objects in our sample are indicated by the vertical yellow lines. {\it c:} $f_{\rm esc}$ as a function of $\beta$. We note that the $f_{\rm esc}$ of the LzLCS represents a direct measure the global escape fraction, while the \fesc\ for our objects is a measure local to the HII regions. Furthermore, $\beta$ for these two samples is measured over a different wavelength range, however the two measures of the UV slope can be compared in a relative sense, such that objects would be `blue' using both measures.}
    \label{fig:fesc-compare}
\end{figure*}

We further explore the physical properties of these systems derived from the SED modeling approach with varying \fesc.  
These properties are summarized in Table~\ref{tab:properties}.  
Unsurprisingly, we find that both objects have minimal dust content, with V-band optical depths ranging from $\tau_{\rm V}=0.01$ to $\tau_{\rm V}=0.02$.  
Furthermore, both objects are relatively low-mass systems, with similar inferred stellar masses of $ \log(M/M_{\odot})=7.8$ and $ \log(M/M_{\odot})=8.0$. 
By assuming a constant star-formation history, we find these two objects span a range of young ages of { 12Myr and 40Myr,} consistent with constraints at the young end of the age distribution among galaxies near this epoch \citep[e.g.,][]{Endsley2021,Whitler2022}.
We constrain the rest-UV sizes of these two galaxies using the NIRCam SW bands and following the procedure outlined in \citet{Chen2022}.  
Both galaxies are consistent with being unresolved, indicating that they are highly compact systems, with effective rest-UV radii below 0.26 kpc (see Figure~\ref{fig:postage}). 
The galaxy properties provided by the SED fitting, in combination with the small size constraints indicates these systems have high SFR surface densities ($\Sigma_{\rm SFR}\equiv \frac{\rm SFR/2}{\pi r_e^2}$). The unresolved nature of these sources means we can only place lower limits on $\Sigma_{\rm SFR}$ of $ >11.1\rm M_{\odot}yr^{-1}kpc^{-2}$ and $ >7.3\rm M_{\odot}yr^{-1}kpc^{-2}$, respectively, for EGS-42501 and EGS-47688, implying their true values may be much higher.

These measured and derived properties allow us to place the two $z>7$ high-\fesc\ candidates in context with samples of known LyC leakers found at lower redshifts, where direct detection of the ionizing radiation escape is possible due to relatively low attenuation from the intergalactic medium (IGM).
In particular, we compare to the Low-redshift Lyman Continuum Survey \citep[LzLCS;][]{Flury2022a, Flury2022b} comprising a sample of objects at $z\sim0.3$, and the Keck Lyman Continuum Survey \citep[KLCS;][]{Steidel2018, Pahl2022} composed of $z\sim3$ Lyman Break Galaxies (LBGs).  
The comparison of our two objects with these lower-redshift samples is provided in Figure~\ref{fig:fesc-compare}.
We note that within these comparisons, the $f_{\rm esc}$ values of the low-redshift samples are global escape fractions based on direct detection, while our results comprise candidates of \fesc\ defined locally within the HII regions of the high-redshift systems.

In Figure~\ref{fig:fesc-compare}(a) we compare the SFRs and sizes of objects within low-redshift samples to our objects, and indicate low-redshift galaxies with confirmed $f_{\rm esc}$.  
It is apparent that galaxies in the low-redshift universe that leak ionizing radiation span a wide range of SFRs, sizes, and thus SFR surface densities.  
While our sources only contain upper limits on $\Sigma_{\rm SFR}$, it is clear that they are consistent with having $\Sigma_{\rm SFR}$ at the high-end of the comparison samples.
Significant escape of ionizing radiation may be expected among high-$\Sigma_{\rm SFR}$ systems, as the intense SFR activity can aid in the creation of channels that allow ionizing radiation to escape \citep[e.g.,][]{Ma2016}. However, high $\Sigma_{\rm SFR}$ does not appear to be a strict requirement for non-zero $f_{\rm esc}$ in these low-redshift systems.
The specific SFR surface density ($\Sigma_{\rm sSFR}\equiv \Sigma_{\rm SFR}/M_*$) provides an alternative metric that has added sensitivity to the depth of the galaxy potential, which may be an important ingredient in regulating the covering fraction, and thus escape of LyC photons \citep[e.g.,][]{Reddy2022}.
Strikingly, our two high-\fesc{} candidates lie at the highest values probed by the low-redshift sample (see Figure~\ref{fig:fesc-compare}(b), supporting this physical picture, however better statistics will be required to further confirm the prevalence of galaxies with high inferred \fesc\ among the high-$\Sigma_{\rm sSFR}$ population.
Finally, we examine $f_{\rm esc}$ as a function of $\beta$ for our two objects in the context of results from the LzLCS sample in Figure~\ref{fig:fesc-compare}(c).
It must be pointed out that due to the available data, both $f_{\rm esc}$ and $\beta$ have different definitions between the two samples, and thus concrete statements regarding their comparison are challenging.
However, it is clear from this figure, and discussed in detail using a slightly alternative approach in \citet{Chisholm2022}, that the abundance of high-$f_{\rm esc}$ objects is qualitatively greater among galaxies with bluer $\beta$, suggesting a potential link between the two quantities.

The low number of such blue galaxies makes it challenging to robustly constrain the abundance of these potentially highly-efficient LyC leakers in the epoch of reionization.  
Despite their apparent low abundance, they may become more common at the faint end of the UV luminosity function and indeed may comprise a significant contribution of the total ionizing photon budget at these earlier epochs.
Future work using \jwst\ will be needed to confirm the presence of this population and constrain their typical properties.
The newly-available spectroscopic capabilities from \jwst\ will further cement the properties of these extreme objects, providing both precise redshifts, as well as constraints of internal ISM properties.

\begin{table*}
\begin{center}
\begin{tabular}{cccccccccc}
\toprule
Object ID & $\log(Z/Z_{\odot})$ & $\tau_{\rm V}$ & $\log(\rm M/M_{\odot})$ & SFR$/M_{\odot}\rm yr^{-1}$&Age [Myr] & \fesc & sSFR [Gyr$^{-1}$] & [OIII]+H$\beta$ EW [\AA] & $\log(U)$ \\
\midrule
 EGS-9466 &   $-1.6^{+0.8}_{-0.4}$  & $0.01^{+0.02}_{-0.01}$ & $7.8^{+0.4}_{-0.1}$ & $5^{+7}_{-3}$ & $12^{+18}_{-7}$ & $0.74^{+0.18}_{-0.44}$ & $79^{+121}_{-46}$ &  $440^{+80}_{-110}$ & $-1.8^{+0.3}_{-0.2}$ \\
 \\
  EGS-16168 &  $-1.5^{+0.7}_{-0.4}$  & $0.02^{+0.02}_{-0.02}$ & $8.0^{+0.4}_{-0.3}$ & $3^{+5}_{-2}$ & $40^{+40}_{-28}$  &$0.49^{+0.40}_{-0.32}$ & $30^{+50}_{-20}$ & $410^{+190}_{-140}$ & $-2.2^{+0.4}_{-0.5}$ \\

 \bottomrule
 \end{tabular}
 \end{center}
\caption{Properties of the extremely blue objects derived using best-fit BEAGLE models allowing for variations in \fesc, as described in \S4. The given uncertainties on these values correspond to the 16th and 84th percentiles of the probability distributions.}
\label{tab:properties}
\end{table*}

%
%
\section{Summary and Outlook}
\label{sec:summary}

In this paper we have described a search for galaxies with 
extremely blue UV continuum slopes ($\beta\simeq -3$) and moderate to weak rest-optical nebular line emission (via color excesses) at $z\simeq 7-11$ in the CEERS NIRCam imaging of the EGS field. Objects that satisfy these criteria are likely to 
be at relatively low metallicity with a large fraction of their ionizing photons escaping from their HII regions. 
We summarise our main conclusions below.

1. We characterize the distribution of UV slopes for a sample of 123 galaxies selected to lie at $z=7-11$, finding a median value of $\beta=-2.0$.  We emphasize that this is merely the observed average, and insight into the intrinsic UV slope distribution (e.g., \citealt{Dunlop2012,Finkelstein2012,Rogers2013,Bouwens2014,Bhatawdekar2021}) requires  detailed simulations that are beyond the scope of this study. Nonetheless, these measurements suggest that our parent sample is comprised of moderately blue UV colors on average, consistent with expectations for these redshifts.

2. Among the sample of high-redshift galaxies, we identify two UV-faint ($\rm M_{\rm UV} = -19.7$ to -19.3) galaxies that exhibit extremely blue UV slopes of $ \beta=-2.9$ and $ \beta=-3.1$ with relatively weak to moderate color excesses. Collectively these observations suggest diminished nebular emission (continuum and lines) and low metallicities. We find that the SEDs cannot be fit well by our fiducial ionization-bounded (\fesc=0) BEAGLE models. We also consider BEAGLE models that approximate the effects of alpha-enhancement (allowing stellar metallicity to be lower than the gas-phase metallicity), finding similarly poor fits. Much better solutions are found when 
comparing to density-bounded BEAGLE models that include a self-consistent treatment of nebular emission while allowing for the escape of ionizing radiation (developed in \citealt{Plat2019}). 
We find that simultaneously reproducing the very blue UV slope and relatively weak color excesses requires density-bounded models  with escape fractions in the range \fesc$ =0.49-0.74$. These objects thus appear to be candidates for significant LyC leakage at $z>7$.

3. We investigate the properties of these extremely blue objects, and find they are low-mass ($ \log(\rm M/M_{\odot})\sim7.8-8.0)$, young (12-40 Myr), and are consistent with being dust-free ($ \tau_{\rm V}\sim0.01-0.02$). 
Each galaxy is unresolved in the UV with NIRCam, with very compact sizes implied ($r_e\lesssim270$ pc).  This allows us to put lower limits on the SFR surface densities, constraining them to large values ($ >(7.3-11)\ \rm M_{\odot}yr^{-1}kpc^{-2}$). These systems thus stand out as dwarf galaxies undergoing recent bursts of star formation, leading to relatively high SFR surface densities.

4. The density-bounded models of these two objects imply they have very low metallicities, with best-fit values of $\log(Z/Z_{\odot})=-1.6$ and $\log(Z/Z_{\odot})=-1.5$.  These low metallicities are implied by the strength of observed photometric excesses resulting from nebular emission lines, such that we infer [OIII]+H$\beta$ EWs inferred from the models, where we obtain values of 440\angstrom{} and 410\AA, for EGS-9466, and EGS-16168, respectively. 

5. We place these two objects (candidates for LyC leakers) in context with low-redshift samples containing galaxies with directly constrained $f_{\rm esc}$. Their SFR surface densities are consistent with those found for leaking systems within low redshift samples, and have comparable $\Sigma_{\rm sSFR}$ to the highest values found in the comparison samples \citep{Pahl2022}.  Finally, we find qualitative agreement with the observed trend between $\beta$ and $f_{\rm esc}$ found at low redshift \citep{Chisholm2022}, such that the strongest leakers typically have very blue UV slopes.  Further work will be required to calibrate these relations at lower redshifts to help interpret samples of LyC leaking candidates  at $z>7$ (such as those in this work) that {\it JWST} will soon discover.

The identification of these extremely blue galaxies (coupled with weak to moderate nebular line emission) hints at the possibility of identifying large numbers of Lyman continuum leaking galaxies in the epoch of reionization, where direct observation of the ionizing continuum is not possible.  Objects with large \fesc{}, and thus low nebular contribution to their SED, offer a unique window on the shape of the stellar continuum (indicating in these objects very low metallicities). Coupling these constraints on the massive star populations with spectroscopic observations that are sensitive to the ISM and gas-phase properties will yield unique insight into early galaxies.

\section*{Acknowledgements}
We thank the anonymous referee for their helpful comments.
DPS acknowledges support from the National Science Foundation through the grant AST-2109066. RE acknowledges funding from NASA JWST/NIRCam contract to the University of Arizona, NAS5-02015. LW acknowledges support from the National Science Foundation Graduate Research Fellowship under Grant No. DGE-2137419.
The authors thank Jacopo Chevallard for use of the BEAGLE tool used for 
much of our SED fitting analysis. 
We thank Gabe Brammer for providing the optical imaging of the EGS as part of CHArGE program. This material is based in part upon High Performance Computing (HPC) resources supported by the University of Arizona TRIF, UITS, and Research, Innovation, and Impact (RII) and maintained by the UArizona Research Technologies department.

\section*{Data Availability}
The data underlying this article will be shared on reasonable request to the corresponding author.

\bibliographystyle{mnras}
\bibliography{main}

\appendix
\section{Posterior distributions of the \fesc{} candidate galaxies.}
\begin{figure*}
    \centering
    \includegraphics[width=0.7\linewidth]{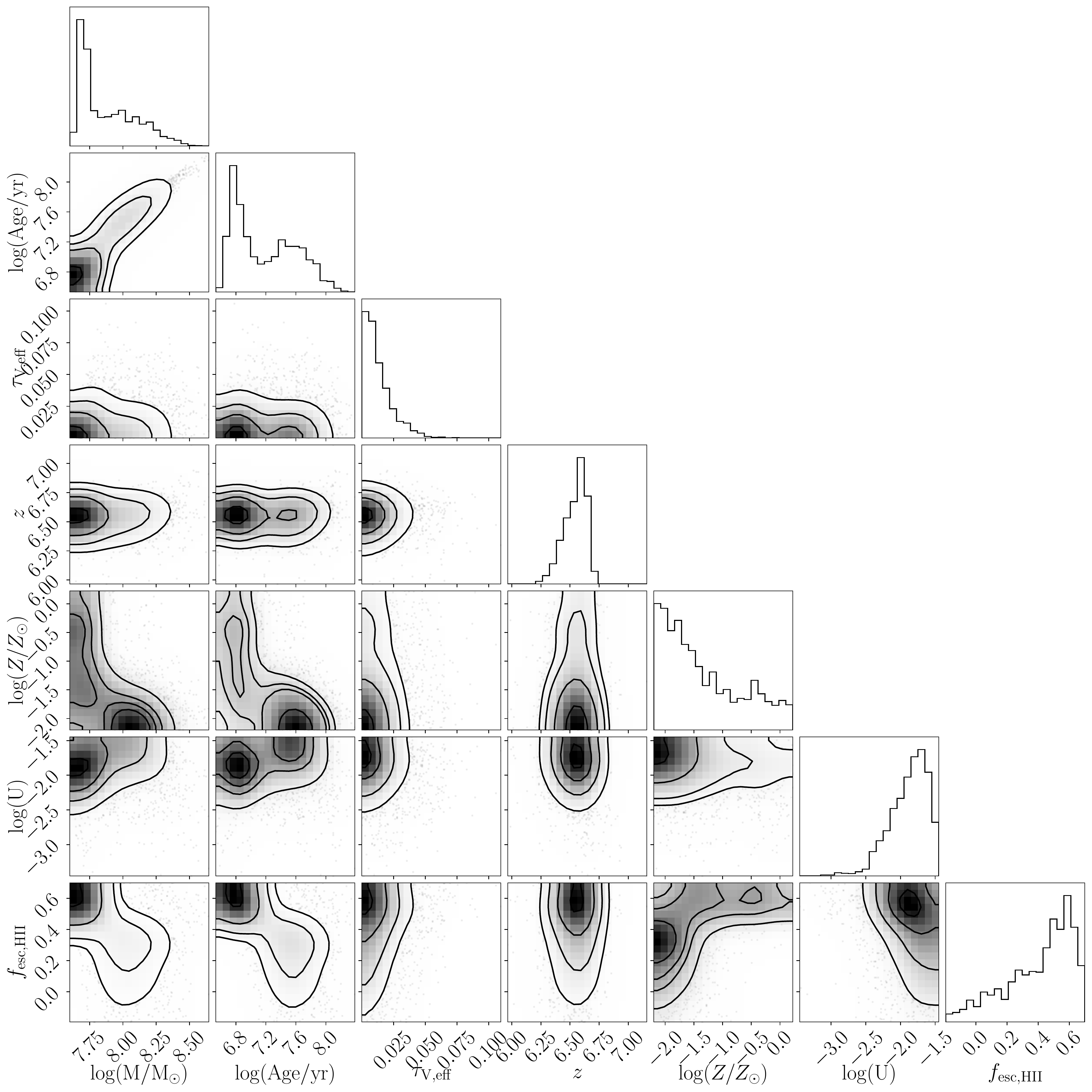}
    \caption{Posterior probability distributions for the parameters inferred for EGS-9466 using the BEAGLE model setup described in \S4.}
    \label{fig:9466_corner}
\end{figure*}

\begin{figure*}
    \centering
    \includegraphics[width=0.7\linewidth]{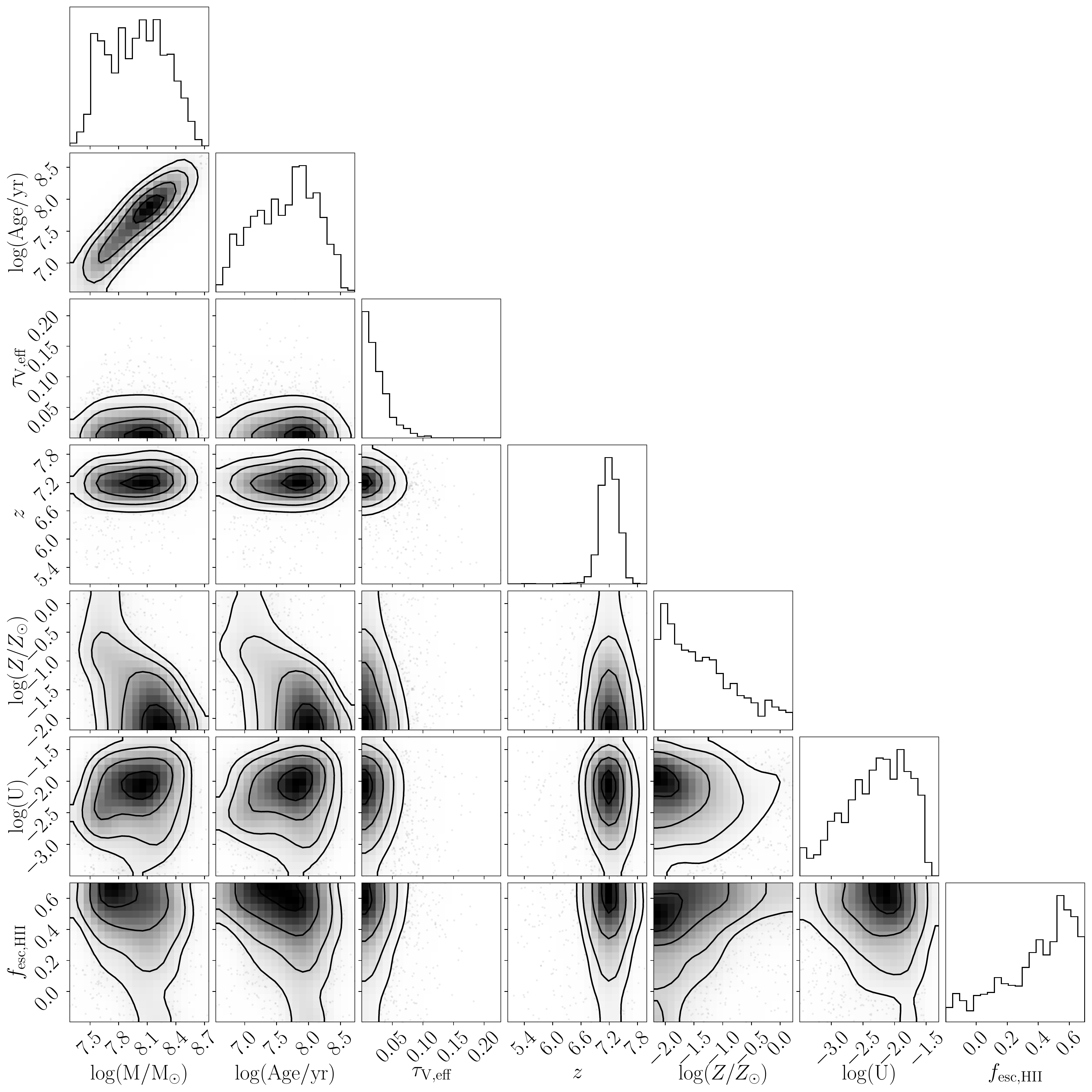}
    \caption{Same as Figure~\ref{fig:9466_corner} but for EGS-16168.}
    \label{fig:16168_corner}
\end{figure*}

 In order to investigate possible degeneracies between free parameters assigned in the SED fitting, we provide corner plots displaying the posterior probability distributions of each fitted parameter, and the covariance between free parameters.
These results from the BEAGLE SED fits are provided in Figures \ref{fig:9466_corner} and \ref{fig:16168_corner} for EGS-9466 and EGS-16168, respectively.

\end{document}